\documentclass[twocolumn,showpacs,preprintnumbers,amsmath,amssymb]{revtex4}

\usepackage{graphicx}
\usepackage{dcolumn}
\usepackage{bm}

\begin{document}

\title{To the concept of the elastic interaction}

\author{Bohdan Lev}

\affiliation{Department of the Theoretical Physics,Institute of
Physics, NAS Ukraine, Prospekt Nauki 46, Kyiv 02022, Ukraine.}

\date{\today}

\begin{abstract}

A general description of elastic matter and the long-range elastic
interaction is propose. The type of the far-field interaction is
determined by the way of breaking in the continuum distribution
of the elastic field produced by topological defects, which can
present isolated inclusions. To provide an adequate description
of the inter-inclusion interaction. Thus we can determine the size
of inclusion as core topological defect in elastic field. In this
description the charge in electrodynamic and mass in gravity
present peculiarity of elastic filed and determined in term this
field. The interaction is a direct and immediate result of the
field deformation. Exist two type interaction. Interaction
through change the ground state of elastic matter and interaction
with the help of the carrier of interaction, which can present as
small changing of deformations.
\end{abstract}
\maketitle

In the general case we have the ground state of the continuum
which can always be described in terms of the elastic field. In
the ground state, this field  has some definite value which does
not depend on the point of the elastic continuum. The different
fluctuation of elastic field which mean value is zero present the
ground state too. To consider the elastic field that determines
this continuum, we have to describe possible deformations of the
distribution of this field which become dependent on the point of
this continuum. The elastic field may be characterized by various
geometrical objects, i.e., a scalar if this field describes the
phase transition in a condensed matter (liquid, liquid crystal,
the wave function in the superconductivity ), a vector potential
in electrodynamics, a second-rank tensor in the general relativity
etc. The basic concept is that in any system with local broken
continuous symmetry exist states in which the elastic variables
describe distortions of spatial configurations of the ground
state. These distortions arise if the continuous symmetry of the
elastic-field distribution is broken in a local area. The first
way to break the continuous symmetry is to introduce a foreign
inclusion in the elastic matter. Then the long-range interaction
between inclusions is determined by the symmetry of the
deformation of the elastic field. This deformations are produced
by the boundary conditions on the inclusions. Then interaction
expressed in terms of the characteristics of this inclusion. In a
system with broken continuous symmetry also can exist defects in
the elastic-field distribution, i.e., the topological defects. In
the general case, each topological defect has some core region,
where the elastic-field distribution is strongly destroyed, and a
far-field region where the elastic variable slowly changes in
space. Both inclusions and topological defects are foreign to the
elastic field and can not be described in terms of this field.
Nevertheless they must have characteristics which influence on the
ground state of the elastic continuum. This characteristics
determine the value and character of the deformation. Then the
amount value of various elastic deformations can be associated
with the effective interaction between inclusions. In other words,
the presence of the field deformation immediately leads to the
interaction. Such interaction did not need the carrier.
Interaction of other type determined by the changes of the global
deformation of elastic field can arise only when inclusions
changing or moving.

Let us consider how the deformation of the elastic-field
distribution can produce interaction of additional inclusions or
topological singularities. We start with the description of a
scalar field with spatially uniform ground state. We have to
describe probable deformations of  the distribution of a scalar
elastic field $\varphi(\overrightarrow{r})$. With an additional
inclusion being introduced in the elastic matter, we can write
the action of this system in the form
\begin{equation}
S=\frac{1}{8\pi}\int\left\{
\left(\overrightarrow{\nabla}\varphi(\overrightarrow{r})\right)^{2}+2\sum_{i}f_{i}\varphi(\overrightarrow{r_{i}})\right\}d\overrightarrow{r}
\end{equation}
where the first term describes the deformation energy of the
elastic field $\varphi$ and the second term is responsible for
the effect of this inclusion, located at the point
$\overrightarrow{r_{i}}$, on the elastic field. Note that the
inclusion must possess properties which influence the elastic
continuum. The first way to describe the interaction of
inclusions is to obtain the change of the deformation energy
produced by this additional inclusions. In the Fourier
presentation $\varphi(\overrightarrow{k})=\frac{1}{(2\pi)}\int
d\overrightarrow{r}\varphi(\overrightarrow{r})exp(-i\overrightarrow{k}\overrightarrow{r})$
we can rewrite the action in the $\overrightarrow{k}$ space in the
form
\begin{eqnarray}
&&S=\frac{1}{8\pi}\frac{1}{(2\pi)^{3}}\\ &&\times\int d\overrightarrow{k}
\left\{\overrightarrow{k}^{2}\varphi^{2}(\overrightarrow{k})+
2\sum_{i}f_{i}\varphi(\overrightarrow{k})exp(i\overrightarrow{k}\overrightarrow{r_{i}})\right\}\nonumber
\end{eqnarray}
In order to find probable configurations of the field $\varphi$ we
have to solve Euler-Lagrange (EL) equations with minimum of
action with regard to the boundary conditions on the inclusion.
This equation is given by
\begin{equation}
\overrightarrow{k}^{2}\varphi(\overrightarrow{k})=-4\pi
\sum_{i}f_{i}exp(-i\overrightarrow{k}\overrightarrow{r_{i}})
\end{equation}
which corresponds to the Poisson equation $\Delta
\varphi=-4\pi\sum_{i}f_{i}$ in the real space. The solution of
this equation yields the field distribution in the form
\begin{equation}
\varphi(\overrightarrow{k})=-4\pi\sum_{i}f_{i}
\frac{exp(-i\overrightarrow{k}\overrightarrow{r_{i}})}{\overrightarrow{k}^{2}}
\end{equation}
With this distribution of the field, we can calculate the
elastic-field deformation energy produced by two inclusions:
\begin{equation}
U_{ij}=\frac{4\pi}{2}\frac{1}{(2\pi)^{3}}\sum_{i,j}f_{i}f_{j}
\int d \overrightarrow{k} \frac {exp
-i\overrightarrow{k}(\overrightarrow{r_{i}}-\overrightarrow{r_{j}})}{\overrightarrow{k^{2}}}
\end{equation}
Having integrated over $\overrightarrow{k}$ we obtain the
interaction energy of two inclusions in the standard form
\begin{equation}
U_{ij}=\sum_{i,j}\frac{f_{i}f_{j}}{\overrightarrow{r_{i}}-\overrightarrow{r_{i}}}
\end{equation}
If we assume that the scalar field is the electrostatic potential
and $f$ is the charge, we obtain the energy of Coulumb-like
charge interaction through the deformation of the electric field.
And, if we assume that the scalar field is the fundamental scalar
field and $b\equiv m$ is the mass, we obtain that this mass
interacts through the deformation of the scalar field according
to the Newton law of repulsive nature. In the two-dimensional
case the interaction energy takes the form \cite{nik}
\begin{equation}
U_{ij}=\sum_{i,j}\frac{f_{i}f_{j}}{2\pi \sigma}ln\frac{d}{r_{0}}
\end{equation}
which is the energy of the interaction of two particles which
distort the interface with the surface tension $\sigma$, where
$r_{0}$ is an arbitrary constant. In order to take into account
the charge dispersion in the matter we have to describe the
distribution of this field in the area of dispersed charges. In
this case, in this the local area the symmetry of the
elastic-field distribution is different and
$\varphi(\overrightarrow{r_{i}})\simeq
\varphi(\overrightarrow{r})+(\overrightarrow{\rho_{i}}
\nabla)\varphi(\overrightarrow{r})$, where
$\overrightarrow{\rho_{i}}$ is the distance to a single charge
and, having introduced the dipole moment
$d_{i}=\sum_{i}b_{i}\rho_{i}$, we find that can find the
dipole-dipole interaction in standard form. The symmetry of
distribution of elastic field which produce two inclusion is
different as symmetry deformation elastic field every inclusion.
Two-particle system selects the solution which corresponds the
position and properties of these inclusions

For a system with broken continuous symmetry, we can consider a
class of defects in the distribution of elastic field that are
called the topological defects. A topological defect can play the
role similar to a particle which changes the elastic field. In
the general case, each topological defect has a core region,
where the elastic-field distribution is strongly destroyed, and a
far-field region where the elastic variable slowly varies in
space. The boundary conditions are then determined by the
conditions on the core of the defect. In this approach we can
find the interaction energy of topological defects. We can start
from action (1) without additional force $f$. The equation of
minimum action in this case is given by the Euler-Lagrange
equation
\begin{equation}
\triangle \varphi(\overrightarrow{r})=0
\end{equation}
This equation has many solutions. The first solution, $\varphi=0$
or $\varphi=const$, is trivial and describes the ground state.
The particular solution of this equation, compatible to the
existence of a topological singularity, may be written as
$\varphi_{n}=(mr^{n}+kr^{-(n+1)})Y_{n}(\theta,\phi)$, and in the
case $n=0$ we have a spherically symmetric solution $ \varphi
=\frac{k}{r}$ where $k$ may be interpreted as the magnitude of
the topological charge. This solution can describe the
singularity in the topological behavior of the scalar field. As a
result, this solution has infinite eigenvalue, however the
interaction energy of such topological charges is finite. A
superposition of two solutions for the scalar field,
$\varphi(\overrightarrow{r})=\varphi^{1}(\overrightarrow{r})+\varphi^{2}(\overrightarrow{r})$,
which produce two topological charges and determine the
interaction energy given by $U_{int}\equiv
E(\varphi^{1}(\overrightarrow{r})+\varphi^{2}(\overrightarrow{r}))-E(\varphi^{1}(\overrightarrow{r}))-E(\varphi^{2}(\overrightarrow{r}))$
yields
\begin{equation}
U_{i j}=\frac{2}{8\pi}\int^{\infty}_{0} (\nabla
\varphi^{1})(\nabla \varphi^{2})d
\overrightarrow{r}=\frac{k_{i}k_{j}}{d}
\end{equation}
where $d$ is the distance between the topological charges. This
formula implies that two topological defects with like signs in
the elastic scalar field repel according to the Coloumb law. In a
more rigorous approach \cite{lupe} the interaction energy of two
defects in the two-dimensional have the same law.

Let us find this solution for a dynamical electromagnetic field.
The action for electrodynamics may be written in the standard
form, i.e.,
\begin{equation}
S=-\frac{1}{16\pi c}\int
\left\{F_{ij}F^{ij}+\frac{16\pi}{c}A_{i}j^{i}\right\}d \Omega
\end{equation}
where the Maxwell stress tensor $F_{ij}=\frac{\partial
A_{i}}{\partial x_{j}}-\frac{\partial A_{j}}{\partial x_{i}}$ is
determined by the vector-potential $\overrightarrow{A}$ and
$\overrightarrow{j}$ is the current of charges. From the minimum
of action, we obtain the field equations as given by
\begin{equation}
\frac{\partial F_{ij}}{\partial x_{j}}=-\frac{4\pi}{c}j_{i}
\end{equation}.
For the  Lorentz gauge condition $\frac{\partial A_{i}}{\partial
x_{}}=0$ in the four-dimensional Euclidean space, the latter
equation came to wave equation for Furier-transformed vector
potential $A_{i}(k,\omega)$ with right term $j_{i}$, whose
solution can write as $A_{i}(k,\omega)=\frac{4\pi c j_{i}exp i
(\overrightarrow{k}\overrightarrow{r}+\omega
t)}{c^{2}k^{2}-\omega^{2}}$. If substitute this solution in the
expression for the action and thus obtain the interaction energy
of different currents in the standard form, i.e.,
\begin{equation}
U_{i,j}=\int d^{4}q j_{i})G^{ij}(q)j_{j}(q)
\end{equation}
where $G(\mathbf{q})$ is the Green function. For an example we
describe the interaction of two charges which move progressively
with the velocity $v$. In this case the charge changes by the law
$e\delta(\mathbf{r}-\mathbf{v}t)$. With regard to the result
\cite{lan}, the Fourier component of the vector potential can be
written as $\varphi_{\mathbf{k}}=4\pi
e\frac{exp(-i(\mathbf{k}\cdot\mathbf{v})t)}{k^{2}-(\frac{\mathbf{k}\cdot\mathbf{v}}{c})^{2}}$
and $A_{\mathbf{k}}=\frac{4\pi e}{c}\frac{\mathbf{v}
exp(-i(\mathbf{k}\cdot\mathbf{v})t)}{k^{2}-(\frac{\mathbf{k}\cdot\mathbf{v}}{c})^{2}}$
Having substituted this vector potential in the expression for
the interaction energy we find that in the case $v=0$ we have the
previous result. In the other case when $j_{0}=e(t)\delta
(\overrightarrow{r})$ yields the interaction energy in the form
$U(\overrightarrow{r}\overrightarrow{r'},t,t')=\frac{e^{2}\delta
(c(t-t')-(|\overrightarrow{r}-\overrightarrow{r'}|))}{|\overrightarrow{r}-\overrightarrow{r'}|}$
that reproduces the standard resultant interaction of variable
charges. This interaction is dynamical and arises if the charge
changes its position or varies in time. The delivery of the field
variation is provided by motion or changes of the global
deformation.

Same result we can obtain if describe the gravitation field and
the appearance the interaction of masses which produce the change
of the geometry of space. The action of the gravitation field and
distributed matter is given by the standard expression
\begin{equation}
S=\frac{c}{16\pi G}\int R \sqrt{-g}d\Omega+\frac{1}{2c}\int T
\sqrt{-g}d \Omega
\end{equation}
where $R$ is the curvature, $T$ is the compression of the
energy-momentum tensor  with the metric tensor $g_{\mu \nu}$,
$\Omega$ is the space-time volume, and $G$ gravitational
constant. Minimization of this action yields the Einstein equation
\begin{equation}
R_{\mu \nu}-\frac{1}{2}g_{\mu \nu}R=\frac{8\pi G}{c^{4}}T_{\mu
\nu}
\end{equation}
For a distributed matter, the of energy-momentum tensor can be
written as $T_{0 0}=-mc^{2}$ and the field equation reduces to
$R_{0 0}=-\frac{4\pi G}{c^{4}}T_{0 0}$. In the linear
approximation we have $\Gamma_{0 0}\simeq -\frac{1}{2}g_{\mu
\mu}\frac{\partial g^{0 0}}{\partial
x_{\mu}}=\frac{1}{c^{2}}\frac{\partial \varphi}{\partial
x_{\mu}}$ that have as result \cite{lan} $R_{0
0}=-\frac{1}{c^{2}}\frac{\partial^{2} \varphi}{\partial
x^{2}_{\mu}}=- \frac{1}{c^{2}}\triangle \varphi$. Thus we obtain
the Poisson equation $\triangle \varphi =4\pi G m$. We substitute
the solution of this equation in the expression for the action
and thus obtain the interaction energy of the distributed matter
in the form of the standard Newton law
\begin{equation}
U =-\frac{G
m_{i}m_{j}}{\overrightarrow{r}_{i}-\overrightarrow{r}_{j}}
\end{equation}
This interaction energy has attractive nature by virtue of the
specifics of gravitation filed which is described by the
second-rank tensor.

In Refs. \cite{inf}, \cite{inf1} the  field equation  in the
empty space was written as
\begin{equation}
R_{\mu \nu}-\frac{1}{2}g_{\mu \nu}R=0
\end{equation}
according to Einstein's statement that the geometry cannot be
mixed with matter. Solution of this equation for a continuum
distribution of matter does not exist. However, there exists a
solution with a singular point in the distribution of matter. In
this presentation we can obtain the same equation for the
gravitational field. The field equation determines the law of
motion in terms of the integral of motion of the surface that
surrounds this singularity. Same presentation of charge it
possible make in electrodynamic where we must solve only equation
for free elastic field
\begin{equation} \frac{\partial
F_{ij}}{\partial x_{j}}=0
\end{equation}.
The charge in electrodynamics, as well as the energy-momentum in
the general relativity, is foreign with respect to the field and
cannot be described in terms of either potentials or the
geometry. They can be only topological singularities in the
distribution of the elastic field. Interaction of these
singularities is governed by the deformation of the elastic field.
In the presentation particle as topological defect is two
preference. First, in this case we have the law of conservation
the topological charge and this not dependent from nature of
elastic field.  Second, we can estimate the size of particle as
the core this defect. The structure this core is structure of
particle. In this picture the charge and gravitational constant
play role the module of tension.

The action for free electrodynamics field can written in the
standard form, i.e.,
\begin{equation}
S=-\frac{1}{16\pi c}\int F_{ij}F^{ij}d \Omega
\end{equation}
The variation this action is follow  \cite{lan}
\begin{equation}
\delta S=-\frac{1}{4\pi c}\int \left\{\frac{\partial
F_{ik}}{\partial x_{k}}\delta A_{i}+\frac{\partial}{\partial x_{k}
}(F_{ik}\delta A_{i})\right\}d \Omega
\end{equation}
Last term go to zero because inasmuch as the surface term and
potential on the surface disappear and this part throw out. But,
if we have topological defect it is not correct.This surface
integral is not zero in  both inside surface of defect and
outside surface the area where exist the electrodynamic field. Can
obtain this non zero term for the defects. For $x_{k}=x_{0}=ct$
the general presentation take the form
\begin{equation}
-\frac{1}{4\pi c}\int \left\{\frac{\partial}{\partial x_{k}
}(F_{ik}\delta A_{i})\right\}d \Omega=-\frac{1}{4\pi c}\int
\left\{\mathbf{\nabla}\varphi \delta \mathbf{A}\right\}d V
\end{equation}
This term can be rewritten in the other form in spherical
coordinates
\begin{equation}
-\frac{1}{4\pi c}\int \left\{\mathbf{\nabla}\varphi \delta
\mathbf{A}\right\}r^{2}dr d \cos\theta d\phi=-\frac{1}{c}\int
\left\{\mathbf{\nabla}\varphi \delta \mathbf{A}\right\}r^{2}dr
\end{equation}
If we take in to account that the solution for defect
$\mathbf{\nabla}\varphi=\frac{k}{r^{2}}\mathbf{r}$, we can
present the last presentation in the form
\begin{equation}
-\frac{1}{c}\int \left\{\mathbf{\nabla}\varphi \delta
\mathbf{A}\right\}r^{2}dr=-\frac{1}{c}\int \left\{k\delta
\mathbf{A}\right\}d\mathbf{r}
\end{equation}
or
\begin{eqnarray}
-\frac{1}{c}\int k\delta \mathbf{A}\frac{d\mathbf{r}}{d
ct}d(ct)&=&-\frac{1}{c}\int k \frac{u_{r}}{c}\delta
\mathbf{A}d(ct)\nonumber\\&=&-\frac{1}{c}\int \left\{\mathbf{j}\delta
\mathbf{A}\right\}d(ct)
\end{eqnarray}
For $x_{k}=r$ we can obtain the other stream of energy through
the surface
\begin{equation}
-\frac{1}{c}\int \left\{\mathbf{\nabla}\varphi \delta
\varphi\right\}r^{2}d ct=-\frac{1}{c}\int k \delta \varphi d ct
\end{equation}
that is density of energy as in previous case which we must make
for deallocation the defect from point zero to infinity. Combining
both obtaining part we can write in four dimensional the
additional part in the action for the electrodynamic field
\begin{equation}
-\frac{1}{c}\int \mathbf{j} \delta \mathbf{A}dV d ct
\end{equation}
The same we cam make from gravitational field \cite{eding}. If we
start from action
\begin{equation}
S=\frac{c^{3}}{16\pi G}\int R \sqrt{-g}d\Omega
\end{equation}
for free gravitation field we can obtain the energy-impulse
tensor as variational parameter which describe peculiarity in
geometry. Every action determine elastic deformation which
produce the topological defect. We can estimate the size of
particle as core this topological defect. This possible make in
two way. The first way is similar. Every core must have the
energy $mc^{2}$, where $m$ is mass of particle or core defect. In
the process of arise two particles must appears two topological
defect with different signs. In this case arise the energy of
interaction between them. We can observe two particle only in the
case when the distance between them will be larger tho sizes of
core. In the condition of equality of energy
$2mc^{2}=\frac{f^{2}}{2R}$ we obtain that the size of core of
size of particle $R=\frac{f^{2}}{4mc^{2}}$. If $f$ is electric
charge we obtained the size of electron, as elementary particle
in the form
\begin{equation}
R=\frac{e^{2}}{4mc^{2}}=\frac{e^{2}\hbar}{4mc^{2}\hbar}=\frac{1}{4}\alpha
\lambda
\end{equation}
$\alpha=\frac{e^{2}}{\hbar c}$ is constant of electric
interaction and $\lambda=\frac{\hbar}{mc}$ is Compton length.The
size of electron is the size of core of topological defect in
electrostatic field and this size is smallest as Compton length in
$\sim 450$ time. In the case gravitation field
\begin{equation}
R==\frac{1}{4}\frac{m^{2}}{m^{2}_{p}}\lambda
\end{equation}
where $m_{p}=(\frac{\hbar c}{G})$ is Plank mass. The elementary
particle in this presentation has not structure. Return to action
of distribution of elastic field. In four dimensional the action
of elastic field in general case can write in the form
\begin{equation}
S=\frac{1}{8\pi}\int\left\{ \left(\nabla \varphi
\right)^{2}-2f\varphi\right\}d\Omega
\end{equation}
where gradient is derivation on the all coordinate and take in to
account the additional part influence the particle or defect on
the distribution of elastic field. Let suppose that elastic field
after broken continuum distribution and formation defect, which
we describe as particle, fluctuated. This fluctuation we can
obtain as elastic field $\psi$ which take in to account the
possible fluctuation of peculiarity of defect, which we can
describe as $\delta f$. After this changing we can rewrite the
action in the new form
\begin{equation}
S=\frac{1}{8\pi}\int\left\{\left(\nabla \psi \right)^{2}- 2\delta
f\psi \right\}d\Omega
\end{equation}
If assume that fluctuations of elastic field interacts with an
additional random field $\delta f$ which involves random vacuum
fluctuations which are not described by the scalar field. Let us
assume that the values of this field are distributed according to
Gaussian law. The average expression for the fluctuations of  the
additional field is given by a continuum integral over the
fluctuations, i.e.,
\begin{equation}
Z\sim \int D\varphi Dh \exp  -\int d\overrightarrow{r}\left(
\frac{1}{2}\left( \nabla \psi \right) ^{2}-2\delta f \psi
-\frac{(\delta f)^{2}}{D^{2}}\right)
\end{equation}
where $D^{2}$ is the dispersion and the last term describes the
energy of fluctuations of the additional random field.
Integrating over all probable configurations of the additional
field $\delta f$ we can obtain the the effective action in the
form
\begin{equation}
S=\frac{1}{8\pi}\int\left\{\left(\nabla \psi
\right)^{2}-D^{2}\psi^{2} \right\}d\Omega
\end{equation}
For elastic filed $\psi$ the equation of field will be meet the
new equation
\begin{equation}
\square \psi=-D^{2}\psi
\end{equation}
For this equation $D^{2}$ must have dimension as
$\frac{1}{L^{2}}$. If take the $L\equiv \lambda $ the equation
for field $\psi$ take the form Laplace equation in four
dimensional for the wave function. The additional deformation of
elastic field which can not formed defect can describe by as
field which can be present as wave of elastic matter. This
presentation not contrary to quantum mechanic because not
determination the position of defect. Formation of this defect
accompany the additional deformation which have the wave nature.


\begin{thebibliography}{99}

\bibitem{lan}  L. D. Landau and E. M. Lifshiz  \newblock{\em Field theory} Nauka, (1973).

\bibitem{nik}  M. G. Nikoladis, A. R.Baush, M. F. Hsu, A. D. Dinsmore,
M. P. Brener, C. Gay and D. A. Weltz \newblock{\em Nature}
\textbf{420} , 299, (2002).

\bibitem{lupe}  T. C. Lubensky, D. Pettey, N. Currier and H. Stark. Phys.Rev.E, \textbf{57}, 610 (1998).

\bibitem{dev}  B. S. DeWitt  \newblock{\em Physical.Report} \textbf{19C}
, 295, (1975)

\bibitem{inf}  L. Infeld ,Rev.Mod.Phys.29,398, (1957).

\bibitem{inf1}  L. Infeld ,Ann.of Physics,6,341, (1959).

\bibitem{eding} A.Eddington,Theory of relativity ,(1938).



\end{thebibliography}
\end{document}